\documentclass{aa}  
\usepackage{graphicx}
\usepackage{txfonts}
\usepackage{natbib}
\pdfoutput=1
\def\asec{\ifmmode ^{\prime\prime}\else$^{\prime\prime}$\fi} 
\def\farcs{\hbox{$.\!\!^{\prime\prime}$}}  
\def\amin{\ifmmode ^{\prime}\else$^{\prime}$\fi} 
\def\farcm{\hbox{$.\mkern-4mu^\prime$}}  
\def\fdeg{\hbox{$.\!\!^{\circ}$}}
\def\grad{$^\circ$}
\def\widerul{\vrule height 2.5ex width 0ex depth 0ex}
\begin{document}
\def\EE#1{\times 10^{#1}}
\def\gcm{\rm ~g~cm^{-3}}
\def\cm3{\rm ~cm^{-3}}
\def\kms{\rm ~km~s^{-1}}
\def\cms{\rm ~cm~s^{-1}}
\def\ergs{\rm ~erg~s^{-1}}
\def\ergsm{\rm ~erg~s^{-1} cm^{-2}}
\def\wl{~\lambda}
\def\wll{~\lambda\lambda}
\def\Nii{M(^{56}{\rm Ni})}
\def\FeI{{\rm Fe\,I}}
\def\FeII{{\rm Fe\,II}}
\def\FeIII{{\rm Fe\,III}}
\def\Niii{M(^{57}{\rm Ni})}
\def\FeIb{{\rm [Fe\,I]}}
\def\FeIIb{{\rm [Fe\,II]}}
\def\FeIIIb{{\rm [Fe\,III]}}
\def\OIb{{\rm [O\,I]}}
\def\OIIb{{\rm [O\,II]}}
\def\OIIIb{{\rm [O\,III]}}
\def\SIIb{{\rm [S\,II]}}
\def\ArIIIb{{\rm [Ar\,III]}}
\def\NiIIb{{\rm [Ni\,II]}}
\def\Msun{~{\rm M}_\odot}
\def\Ti44{M(^{44}{\rm Ti})}
\def\MZA{M_{\rm ZAMS}}
\def\mum{\mu{\rm m}}
\def\snr{SNR~0540-69.3}
\def\psr{PSR~B0540-69.3}

\def\lsim{\!\!\!\phantom{\le}\smash{\buildrel{}\over
  {\lower2.5dd\hbox{$\buildrel{\lower2dd\hbox{$\displaystyle<$}}\over
                               \sim$}}}\,\,}
\def\gsim{\!\!\!\phantom{\ge}\smash{\buildrel{}\over
  {\lower2.5dd\hbox{$\buildrel{\lower2dd\hbox{$\displaystyle>$}}\over
                               \sim$}}}\,\,}

   \title{Optical 
   identification of the 3C~58 pulsar wind nebula}

\author {Yu. A. Shibanov\inst{1}
\and N. Lundqvist\inst{2}
\and P. Lundqvist\inst{2}
\and J. Sollerman\inst{2,3}
\and D. Zyuzin\inst{4}
}

\institute{Ioffe Physical Technical Institute, Politekhnicheskaya 26,
St. Petersburg, 194021, Russia
\and Stockholm Observatory, AlbaNova Science Center, Department
of Astronomy, SE-106 91 Stockholm, Sweden
\and Dark Cosmology Center, Copenhagen, Denmark
\and  Academical Physical Technical University, Khlopina 8, St. Petersburg, 195220, Russia  
}



 
  \abstract
{}
    {3C~58 is a Crab-like supernova remnant containing  the young pulsar PSR J0295+6449,  
    which powers a radio plerion and a compact torus-like pulsar wind nebula visible 
    in X-rays. We  have  performed a deep optical imaging of the 3C~58 field 
    with the Nordic  Optical Telescope in the $B$ and $V$ bands to detect 
    the optical counterpart of the pulsar and its wind nebula.} 
   {We analyzed our data together with the archival data obtained with the  Chandra/ACIS and HRC 
   in X-rays and with the Spitzer/IRAC in the mid-infrared.}
   {We detect a faint extended elliptical optical object with $B$=24\fm06$\pm$0.08 and 
   $V$=23\fm11$\pm$0.04 whose peak brightness and center position are consistent at 
   the sub-arcsecond level with the position of the pulsar. The morphology and orientation 
   of the object are  in excellent agreement with the  torus region of the pulsar wind nebula,  
   seen almost edge on in the X-rays, although its extension is only about a half of that 
   in X-rays.  This suggests that in the optical we see only the brightest central part of 
   the  torus with the pulsar.  The object is also practically identical to the  
   counterpart of the torus region recently detected in the mid-infrared. We estimate that 
   the contribution of the pulsar to the observed optical flux is $\la$ 10\%.  
   Combinig the optical/mid-infrared fluxes  and the   X-ray power-law spectrum extracted 
   from the spatial region constrained by  the optical/infrared source extent we compile 
   a tentative multi-wavelength spectrum  of the central part of the nebula. Within the uncertainties 
   of the interstellar extinction towards 3C~58 it is reminiscent  of either the Crab or  
   PSR B0540-69 pulsar wind nebula spectra.}  
  {The properties of the  detected object strongly suggest 
   it to be the optical counterpart of the 3C~58 pulsar $+$ its wind nebula system. This makes 
   3C~58 the third  member of such a class of the torus-like pulsar$+$nebula   systems  identified  
   in the optical and mid-infrared.}

\titlerunning{Optical identification of the 3C~58 PWN}
\authorrunning{Yu. Shibanov et al.}

   \maketitle
%
\section{Introduction}
The ``Crab-like" supernova remnant (SNR) 3C 58
was first discovered in the radio \citep{ws71}  
and then identified in the optical by observations in
H${\alpha}$  \citep{vdb78}.    Just as the Crab nebula, 
this SNR has a plerionic, or filled center. It shows a filamentary
structure in the H${\alpha}$ emission and in the radio with a flat radio
spectrum of synchrotron origin. The distance, $d=3.2$~kpc, 
and size, $\sim6\amin\times9\amin$ of 3C 58 is also similar to that of the Crab SNR.
What makes SNR 3C~58 particularly interesting is its possible association
\citep{sg02}  with the historical supernova SN 1181. 
The deduced age of $\sim830$~yr is consistent with the remnant being of roughly
the same size as the Crab nebula, but this has been questioned on another grounds (see below).

It was long suspected  that 3C 58 contains a pulsar in its center \citep{bhs82},   
and after several years of searching in X-rays and radio, the pulsar J0205+6449 
with a period of $P=65.68$~ms was discovered with the Chandra X-ray observatory 
at the center of the SNR plerion \citep{mu02}. 
It was directly confirmed also in the radio \citep{mal01, cam02, mal03}.  
Deeper Chandra exposures revealed a Crab-like X-ray torus+jet pulsar wind nebula (PWN) in the center 
of the pulsar powered plerion \citep{sla04}.

One of the puzzles of the 3C 58 system is that, unlike for the 
Crab, the characteristic age of the 3C 58 pulsar, 
$\tau=P/2\dot{P}\approx5400$ yr, is 
considerably higher than the historical age of 830 yr. This can perhaps be
understood by assuming a larger initial spin period of the neutron star (NS). 
However, the factor of $\sim2$ lower radial velocities of the optical
\citep{fes83,fes07}  and radio \citep{beit01,beit07} filaments than for the Crab, 
i.e., $\la 900$ km~s$^{-1}$, either would require a larger actual age of 3C 58, 
a weaker SN explosion, or   a substantial deceleration of the remnant 
\citep{chev04}. Furthermore, the radio emission of 3C 58 is about 10 times
fainter than for the Crab while its X-ray nebula is about 2000 times weaker. 
The latter is surprising  and  does not match the present spin-down power of the pulsar: 
the spindown luminosity of PSR J0205+6449 is $\dot{E}=2.7\times
10^{37}$ erg~s$^{-1}$, and thus 5\%  of that  of the Crab 
 (this is actually the third largest spin-down power for all known 
pulsars in the Galaxy).  Deeper studies of  PSR J0205+6449 and
its PWN at different wavelengths may help to establish a link   
between the  properties of this Crab-like SNR and  its pulsar activity.

Multiwavelength studies of PSR~J0205+6449 are still rare.
Its radio luminosity  is lower than for 99\%~of all known pulsars and 
actually the lowest among known young pulsars 
\citep{cam02}. The X-ray properties of the nonthermal
emission of the pulsar (spectral slope,  pulse shape)   
are similar to those of the Crab.  However, an important difference is the presence 
of a  thermal blackbody-like component from the surface of the cooling NS
with a temperature of about $10^6$~K \citep{sla04}. 
Assuming an age of 830 yr, this temperature falls far below the predictions of the standard NS cooling 
theories and could suggests an enhanced cooling provided by the presence 
of exotic matter (like pion condensates) in the NS interior.  
The spectral index of the X-ray synchrotron emission from the  
torus-like PWN is compatible with  that of the Crab, but 
the luminosity, $\sim5.3\times 10^{33}$ erg~s$^{-1}$,  is about 4000 times smaller. 
The size of the plerion core visible in X-rays, $\la$0.55 pc, 
is only a factor of  3  smaller than that of the Crab. 

There have been few studies of the 3C 58 field in the optical range,
and we are not aware of any deep broadband imaging of the remnant.
Such studies, carried out 
to reveal the continuum emission from the pulsar and the PWN and 
thus to constrain the properties of the 
multiwavelength spectral energy distribution, have been performed for the Crab 
and its twin in the LMC, PSR B0540-69  \citep[e.g.,][]{sol03,ser04}.  
The latter two systems show a strong spectral break between the optical and X-rays
leading to a smaller optical flux than would be expected from a
simple extrapolation of the observed X-ray spectrum toward the optical range. 
 The detection of the 3C~58 PWN in the mid-infrared (mid-IR) recently reported by \citet{sla08} 
also suggests a break between the infrared and X-rays, and perhaps multiple breaks. 
Optical studies are necessery to constrain the position and the number of the breaks.

Here we report on deep optical imaging of the central part of 3C 58 
in the $B$ and $V$ bands with the Nordic Optical Telescope (NOT) on La Palma. 
These observations allowed us to find a likely candidate for the optical counterpart 
of the pulsar+PWN system. We compare our data with X-ray and mid-IR   
data retrieved from the Chandra and Spitzer archives. 
The observations and data reduction are described in Sect.~2. In Sect.~3 we present 
the results which are finally discussed in Sect.~4.           
\section{Observations and data reduction}
\subsection{Observations}
The field  of 3C 58  was observed  on the night of 
October 22-23 2006 with the Andalucia Faint Object Spectrograph and Camera
(ALFOSC\footnote{www.not.iac.es/instruments/alfosc}) 
 at the NOT during a service mode observation (program P34-026).  
ALFOSC  was equipped with a $2048 \times 2048$ pixel CCD  providing a field-of-view of
$6^{\prime}.5 \times 6^{\prime}.5$ with the pixel size on the sky of 0\farcs19.  
The center of 3C 58 was imaged in the $B$- and $V$-band filters
which have throughputs similar to the Bessel system. 
Sets of 10~min dithered exposures were obtained in each of the filters at an  
airmass varied  in a narrow range of $\sim$1.24--1.35.   
The total  exposure time  was 5400~s  in the $B$ and 6600~s in the $V$ bands.  
The PG0231+051 standard field \citep{lan92} 
was used for the photometric calibration, and was observed immediately 
after the 3C~58 field at a similar airmass.
The observing conditions were rather
stable with the seeing varying from $\sim0\farcs6$ to $\sim0\farcs9$ and with mean values 
 of $\sim0\farcs84$ and $\sim0\farcs71$ in the $B$ and $V$ bands, 
 respectively. Standard data reduction  including bias subtraction, flat-fielding and 
cosmic ray rejection  was performed making use of the {\tt IRAF} 
{\it ccdred} package and {\it crrej} task. The individual images in each band 
were aligned and combined with the {\it imcombine} task.
\subsection{Astrometric referencing} 
The astrometric referencing of the NOT images was done making use of 
the USNO-B1.0 catalog \citep{Monet} and the {\tt IRAF} tasks {\tt ccmap/cctran}. 
We chose the resulting $B$-band image as a primary for the referencing since this 
image have fewer saturated stars in the field.
To minimize any geometrical distortion effects  we used the positions of eleven unsaturated reference 
stars\footnote{USNO-B1.0 stars used for the  astrometric transformation of 
the 3C 58 field:  1548-0060227, 1548-0060191, 1548-0060258, 1548-0060348,    
 1548-0060300, 1548-0060306  1548-0060264,  1548-0060155, 1548-0060199, 1547-0060002, 
 1547-0060015.} 
 located within an arcminute from the center of the image.  
 The nominal catalog  position errors of the selected stars are less 
 than 0\farcs1. These stars show no 
considerable proper motions within a few mas yr$^{-1}$.           
Formal {\it rms}  errors of the astrometric fit for the RA and Dec were 
$\approx0\farcs059$ and $\approx0\farcs062$, respectively, and maximal 
residuals of any reference star were $\la0\farcs1$ for both coordinates.   
Using a set of stars with good Point Spread Functions (PSFs) 
the $V$ band image was aligned to the $B$ band image with an accuracy 
of better than $\approx0.1$ pixel, or $\approx0\farcs019$. 
Combining the errors, a conservative estimate of our 1$\sigma$ astrometric 
referencing  accuracy is $\la0\farcs1$ in both RA and Dec for both bands. 
\begin{figure*}[t]
\setlength{\unitlength}{1mm}
\begin{center}
\includegraphics[width=17cm, clip]{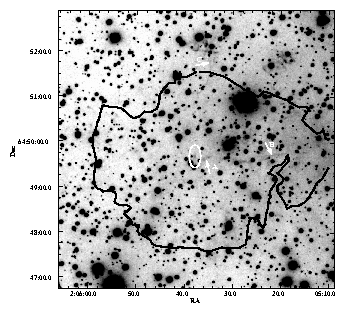}
\end{center}
\caption[h]{Overview ($\sim$6\amin$\times$6\amin) of the center of the
supernova remnant 3C 58 in the $B$-band, as obtained with NOT/ALFOSC. 
An outer contour of X-ray emission of the 3C~58 plerion is overlaid from the Chandra/ACIS-S image.  
A bold ellipse in the plerion center marks the position where 
the pulsar J0205+6449 powering the nebula is located. The central region is 
enlarged in Fig.~\ref{f:B-V-hrc-ac-sp-vla}. White arrows marked ``A", ``B" and
``C" point at some filaments belonging to the remnant. Filament A is discussed 
further in Fig.~\ref{f:B-V-hrc-ac-sp-vla}.         
}
\label{f:3c58}
\end{figure*}    
\subsection{ Photometric  calibration} 
The observing night was photometric.    
The photometric calibration was carried out 
using six standard stars from the Landolt field PG0231+051 
\citep{lan92} and the {\tt IRAF} packages {\tt daophot} and {\tt photcal}. 
The small variations  of the airmass during our 
observations did not allow us to estimate the atmospheric extinction coefficients  from our own data.
We therefore fixed the extinction coefficients at their mean values adopted 
from the NOT homepage: $k_B$=0$\fm$22 and $k_V$=0$\fm$12.  Since the target and the standard stars
had small airmass differences the uncertainties in the airmass correction
has a negligible effect on our photometry. As a result of the photometric fit,  
we obtained the following zeropoints $B_{ZP}$=25\fm62$\pm$0.01 and $V_{ZP}$=25\fm51$\pm$0.01,
and color terms $BV_b$=0\fm036$\pm$0.011 and $BV_v=-0\fm068\pm$0.014. 
\section{Results} 
\subsection{Overview of the 3C 58 field} 
The NOT/ALFOSC overview of the 3C 58 field in the $B$ band is shown in Fig.~\ref{f:3c58}.  
The field is crowded with stars which makes it difficult to immediately 
identify any optical counterpart of the plerion's X-ray emission, whose 
outer contour  from an archival Chandra/ACIS-S image\footnote{The X-ray data were 
retrieved from the Chandra archive (Obs ID 4382,2003-04-23, 170 ks exposure, 
PI P. Slane).} is overlaid.  Nevertheless, one can resolve several bright optical 
filaments likely associated with the SNR, e.g., a long filament extending
from the center of the nebula in the E-W direction.   
The brightness of the extended optical emission increases toward the 
X-ray tail west of the plerion. 
The same is seen in our $V$ image and in a red Palomar plate \citep{vdb78,fes83}.    
The long E-W filament also correlates with this tail and it is 
clearly detected in narrow band H$\alpha$ and [O~III] images \citep{fes07}.   
Optical spectra of some central parts of 
the filament have been obtained by \citet{fes83} and reveal a radial velocity 
of up to 900~km~s$^{-1}$ confirming that it belongs to the SNR.       
\begin{figure*}[t]
\setlength{\unitlength}{1mm}
\includegraphics[width=18.0cm, clip]{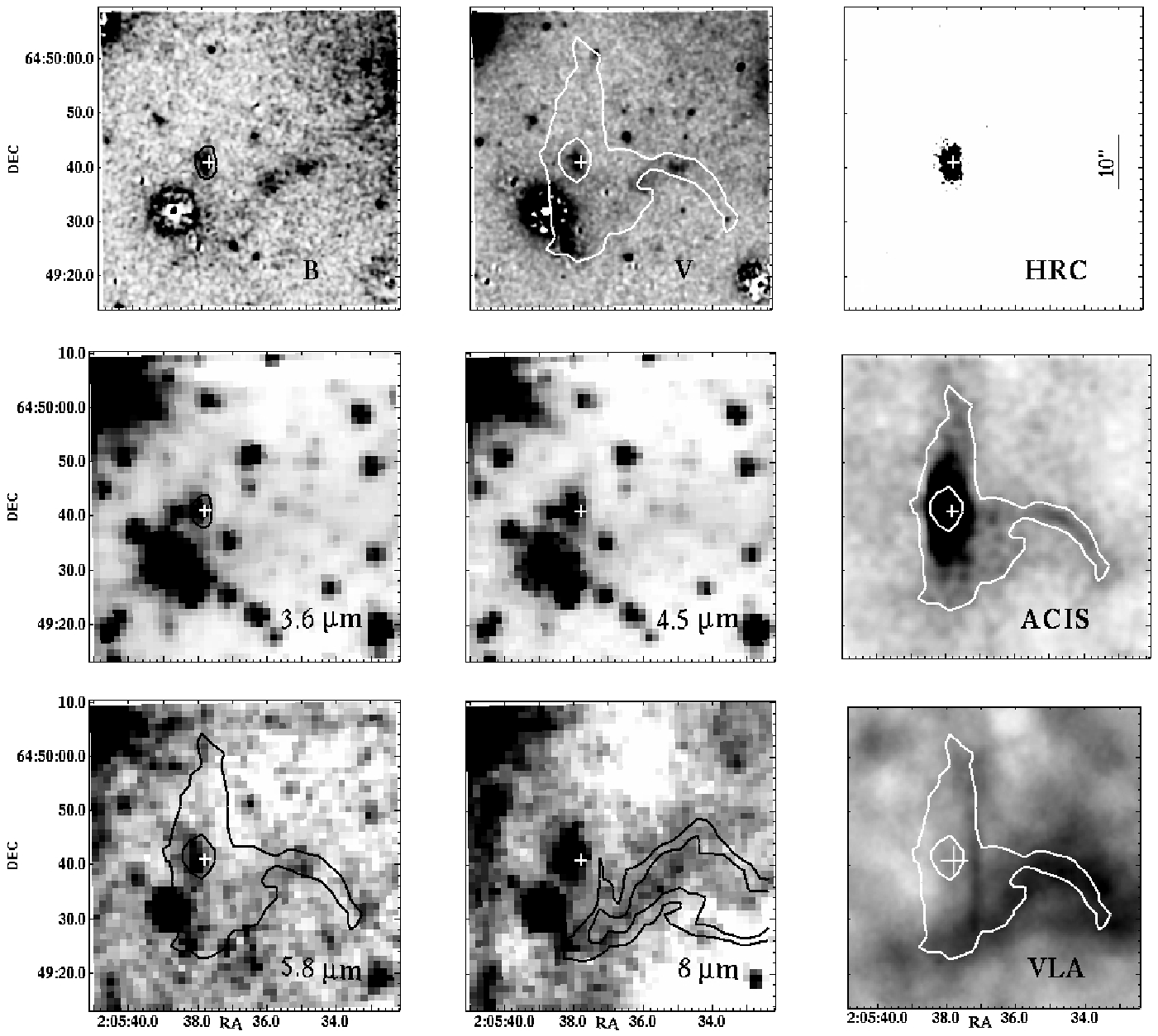} 
\caption[h]{Central $\sim$1\farcm$1\times$1\farcm1~part of 
3C~58 containing the pulsar and the PWN. {\sl Top-Left} and {\sl Top-Middle} 
panels are NOT/ALFOSC $B$ and $V$ band images, respectively, where unsaturated 
field stars have been subtracted to better reveal the suggested pulsar/PWN 
optical counterpart. Other panels show images of the same field but in X-rays, 
mid-IR and radio, as obtained with the Chandra/HRC and Chandra/ACIS, 
Spitzer/IRAC and VLA, respectively, as indicated in the images.       
The optical and X-ray images have been smoothed with a Gaussian kernel 
of three pixels. The X-ray contours outlining the boundary and 
PWN brightness distribution in X-rays on the HRC and ACIS images 
are overlaid on the B, 3.6 $\mu$m and on the V, 5.8 $\mu$m and VLA images, 
respectively. The VLA contours are shown in the 8 $\mu$m image.  
The white cross ("$+$") in all panels marks the Chandra/HRC position of the 
pulsar \citep{mu02}. 
Filament ``A" in Fig.~\ref{f:3c58}
is clearly seen in both $B$ and $V$ to the west of the white cross.         
}
\label{f:B-V-hrc-ac-sp-vla}
\end{figure*}    
Whether the other small scale optical $BV$ filaments and knots are associated 
with similar structures detected in the radio, narrow band optical,  
and/or X-rays is outside the scope of this paper.       
Here we concentrate on the faint extended emission located in the center 
of the image near/within the region marked in Fig.~\ref{f:3c58} 
by a bold ellipse that contains the pulsar.   
\subsection{Detection of the pulsar/PWN optical counterpart}   
The region containing the pulsar is enlarged in
Fig.~\ref{f:B-V-hrc-ac-sp-vla}, where we compare our optical 
$B$ and $V$ images with available archival images obtained in other
spectral domains: Chandra/ACIS-S and HRC-S\footnote{The X-ray data were 
retrieved from the Chandra archive (Obs ID 1848, 2002-01-09, 33.5 ks exposure, 
PI S. Murray).} X-ray images, mid-infrared images obtained with
Spitzer/IRAC\footnote{Program ID 3647, exposure  5.4 ks, PI P. Slane}, and 
a 1.4 GHz VLA radio image \citep{beit07}. 
The  unsaturated stars closest to the pulsar 
were subtracted in the optical images.
The dynamical range of the ACIS image was changed compared to the 
contour shown in Fig.~\ref{f:3c58} to reveal the
structure of the torus-like PWN with its possible western jet 
(assuming the torus is seen edge-on). 

In both optical bands ({\sl Top Left and Middle} panels 
of Fig.~\ref{f:B-V-hrc-ac-sp-vla}) we detect a faint extended, elliptical 
structure at the pulsar position and we have marked this by a white plus sign. 
Its morphology matches the structure seen in the HRC image, which
reveals only the brightest part of the PWN in X-rays. 
Both the extent of this elliptical optical source and the orientation of its
major axis are similar to what is found in X-rays. 
This suggests that the detected source is the optical 
counterpart of the PWN of PSR J0295+6449, with perhaps also a contribution
from the pulsar itself. 
The source is also clearly visible in all four IRAC mid-IR bands 
at the same position and with the same elliptical morphology  
and orientation. An 8~$\mu$m image of this field 
was presented by \citet{slane07}  and  \citet{sla08}.  
The astrometric accuracy of the pipeline produced mid-IR  
post-BCD (Basic Calibrated Data) images shown in 
Fig.~\ref{f:B-V-hrc-ac-sp-vla} is $\sim$0\farcs2, which allows us to 
state that structures we see in the optical and IR represent the same 
source. The mid-IR counterpart appears to be slightly blurred 
in comparison with the optical structures. This is because
of the lower spatial resolution of the IRAC images (pixel size of 1\farcs2)
compared to that in the optical (pixel size of 0\farcs19).  
\begin{figure*}[t]
\setlength{\unitlength}{1mm}
\begin{picture}(180,68)(0,0)
\put(0,-2){\includegraphics[width=5.2cm,clip]{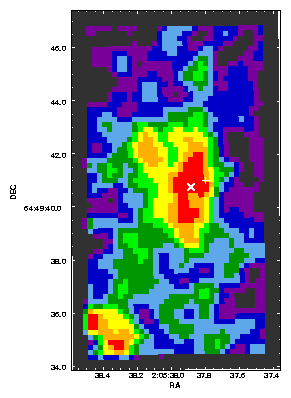}}
\put(62,-2){\includegraphics[width=5.2cm,clip]{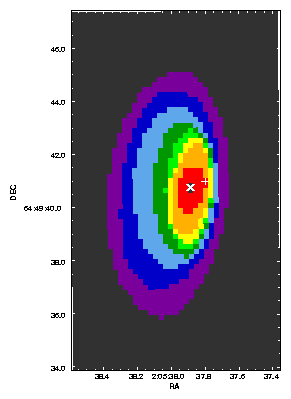}}
\put(123,-2){\includegraphics[width=5.2cm,clip]{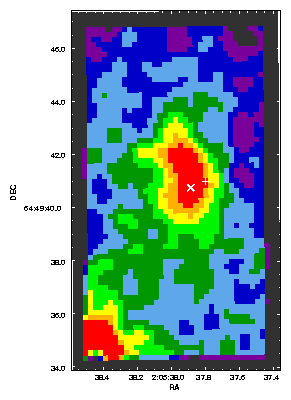}}
\end{picture}
\caption[h]{The 13\asec$\times$6\asec~central region of 3C 58 with the suggested 
optical counterpart of the pulsar/PWN system in the $B$  ({\sl Left}) 
and $V$ ({\sl Right}) bands and results of the $B$ band surface brightness 
fit with elliptical isophotes ({\sl Middle}). As in Fig.~\ref{f:B-V-hrc-ac-sp-vla}, 
the white plus sign marks the HRC-S X-ray position of the pulsar 
while "$\times$" shows the position of the center of the brightest 
part of the optical source obtained from the fit. The difference between the plus and $\times$ is     
$\approx$0\farcs6, which is about twice smaller than the 1-$\sigma$  uncertainty of the pulsar X-ray position 
combined from $\sim$1\asec~HRC  pointing accuracy  and  $\sim$0\farcs1  uncertainty 
of the astrometric referencing of the optical images.   
The $B$ and $V$ images were smoothed with a Gaussian kernel of three pixels. 
The bright source at the bottom left corner of the $B$ and $V$ images is the 
remains of a poorly subtracted field star.            
}
\label{f:ellip}
\end{figure*}    

In the optical we also see two knots on an extended filament west 
of the pulsar (called Filament ``A" in Fig.~\ref{f:3c58}). They coincide 
spatially with an elongated X-ray structure 
seen in the ACIS image that has been interpreted as a possible jet emanating
from the pulsar \citep{sla02}. However, in the optical images we 
do not see the ``beginning" of the jet. The same is true for the 
mid-IR where we see both faint emission from the filament and
the knots, in particular the western-most of the two optical knots. 
The filament, and especially the knots, are best seen at the longest 
IR wavelengths. These structures seen in the optical and mid-IR  
are likely parts of a long curved SNR filament clearly visible in the 
VLA radio image ({\sl Bottom-Right} panel of Fig.~\ref{f:B-V-hrc-ac-sp-vla}).   
To emphasize this, we have overlaid the radio contours on the 8 $\mu$m image. 
The 8 $\mu$m emission traces the radio structure well. At shorter wavelengths,
and in particular in the optical, the full structure of the radio emission
becomes less obvious. That the emission becomes stronger, is less clumpy, and 
traces the radio structure progressively better as the wavelength increases 
from the $B$-band to 8 $\mu$m, makes us confident that we have indeed identified the 
optical/IR  counterpart to the radio filament. 
The X-ray contours overlaid on the radio image also show that at least the 
distant part of the elongated X-ray structure is likely of the same origin 
as the radio filament. This rules out the interpretation of this part of the structure 
as the pulsar jet.

To compare the suggested optical counterpart to the X-ray PWN torus structure,
we have fitted the spatial intensity distribution  with a simple two-dimensional model 
(using the {\tt IRAF} {\tt isophotes} package) accounting for the brightness 
distribution by elliptical isophotes and taking into account the background 
level. The same  was done by \cite{mu02} for the extended part of the X-ray source 
in the HRC image, using {\tt CIAO} {\tt Sherpa} tools.   
The results of the spatial fits are presented in Fig.~\ref{f:ellip} and  in 
Table~\ref{t:ell}. 
\begin{table}[h]
\caption{Parameters of the elliptical fit to the surface brightness of the 
PSR J0295+6449$+$PWN optical counterpart ordered by the major ellipse axis 
length (c.f., {\sl Middle} panel of Fig.~\ref{f:ellip}).}
\begin{tabular}{lllcll}
\hline\hline 
 Center$^a$ & Center$^a$  &  Major          & Ellip-      & PA$^c$       & B-band \\  
  RA	    & DEC	  &   axis          & -ticity$^b$ & N to E   & flux$^d$  \\  
  	    &  	          &  length               &             &    &   \\  
  02:05:* & 64:49:*   &  arcsec       &             & degrees  & \% \\  
\hline                                                                                       
 37.88      & 40.7	  &   0.84	    & 0.28	  & $-$7.29    & 3.1(0.8)   \\ 
 37.88      & 40.7	  &   1.26	    & 0.28	  & $-$7.29    & 7.6(1.2)  \\ 
 37.88      & 40.7	  &   1.9	    & 0.28	  & $-$7.29    & 17.4(1.7)  \\
 37.88      & 40.7	  &   2.86	    & 0.56	  & $-$16.13   & 22.9(2.2)  \\ 
 37.88      & 40.7	  &   4.28	    & 0.62	  & $-$3.68    & 40.9(3.1)  \\
 38.00      & 40.8	  &   6.42	    & 0.48	  & $-$3.68    & 86.3(6.6)   \\
 38.00      & 40.5	  &   9.62	    & 0.59	  & $-$5.40    & 100  \\ 
\hline
\end{tabular}
\label{t:ell}
\begin{tabular}{ll}
$^a$~coordinates of centers of the ellipses (J2000);& \\
$^b$~defined as $1-l_{min}/l_{max}$, where $l_{min}$ and $l_{max}$ are the minor and  & \\
major ellipse axes lengths, respectively; & \\
$^c$~ positional angle of the major axis;& \\ 
$^d$~ flux from the elliptical aperture normalized to the flux from the  &  \\
largest aperture in this set. Numbers in brackets are 1$\sigma$ uncertainties.  &\\
\end{tabular}
\end{table}
 
The fit shows that the brightest part of the optical nebula, 
within 1\asec--2\asec~ of the center, has almost a circular shape and emits 
only $\la 10\%$ of the total flux. If our identification is correct, this 
can be considered as an upper limit of the pulsar contribution to the total 
pulsar$+$PWN optical flux. This is consistent with what was found for two known  
pulsar$+$PWN systems in the optical range \citep{ser04}. The coordinates of the center defined from the 
fit are  RA=02:05:37.88 and DEC=64:49:40.7 (marked by "$\times$" in Fig.~\ref{f:ellip}). 
This is in agreement with the pulsar X-ray coordinates (marked by plus),  
RA=02:05:37.8 and DEC=64:49:41 \citep{mu02}, when accounting for our astrometrical uncertainty, 
$\la 0\farcs1$, and the typical HRC pointing uncertainty, $\la 1\asec$. 
The ellipticity increases up to $\sim$ 0.5--0.6 for the outer optical nebula 
regions, and the major axis is slightly tilted towards the North-West 
by a few degrees, which is also in a good agreement with what was found in X-rays 
\citep{mu02, sla02}.   In addition, the optical nebula is elongated in the north-south direction 
almost symmetrically around its center with a total length of $\sim$9\farcs6, 
which agrees with the $\sim$10\asec~extent of the brightest part of the nebula 
in the HRC-S data. Finally, assuming that we see a tilted  torus-like shaped 
nebula, our elliptical fit suggests  that the angle of the torus symmetry 
axis to the sky plane is in the range of 53\grad - 66\grad. This  is close to  
the value of $\sim$70\grad\ estimated from the X-ray data by \citet{sla02}.

\begin{figure}[t]
\includegraphics[width=8.0cm, clip]{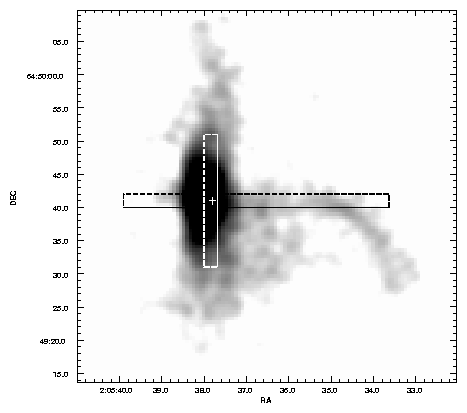}
\caption[h]{ACIS-S image with positions of two slices of 40\asec~({\sl black})
and 20\asec~({\sl white}) lengths and of 2\asec~widths used for the comparison 
of the X-ray and optical spatial profiles of the pulsar$+$PWN system  
in Fig.~\ref{f:prof}.}
\label{f:sl-pos}
\end{figure}

\begin{figure*}[t]
\setlength{\unitlength}{1mm}
\begin{picture}(180,80)(0,0)
\put(10,0){\includegraphics[width=5.8cm,clip]{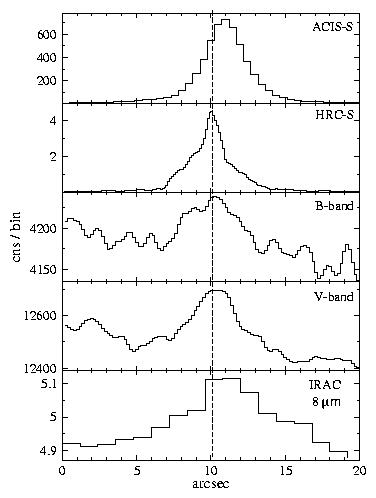}}
\put(85,0){\includegraphics[width=5.8cm,clip]{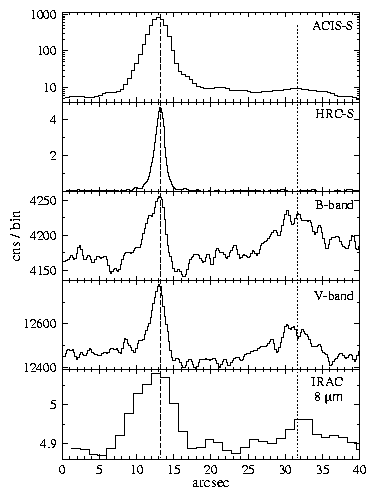}}
\end{picture}
\caption[h]{Spatial profiles of the pulsar/PWN system in the optical,
mid-infrared and X-rays along the north-south ({\sl Left}) and east-west 
({\sl Right}) slices shown in Fig.~\ref{f:sl-pos} obtained from different 
images, as indicated in the plot. The coordinate origins of the horizontal 
axes correspond to Dec=64:49:31:00 and RA=02:05:39:90 for the {\sl Left}
and {\sl Right} plots, respectively. The images were smoothed with a 3-pixel 
Gaussian kernel. Dashed vertical lines indicate the HRC-S position of the 
pulsar. Dotted line shows the position of a bright jet-like structure to the
west, before it bends to the south. The ACIS-S profile in the right panel is 
in a {\sl log} scale to better resolve this structure in X-rays.   
}
\label{f:prof}
\end{figure*}

To bring out more details of the structure of the nebula in the optical/mid-IR 
and in X-rays, we considered also 1D-spatial profiles of the presented 
images along two slices shown in Fig.~\ref{f:sl-pos}. The slices 
are 2\asec~wide with PA 0\grad~and 90\grad, both centered on the HRC-S pulsar 
position. The first slice is 20\asec~long to include the most north-south extent 
of the nebula as seen in the ACIS-S image and was placed symmetrically over 
the center of the nebula. The second slice is 40\asec~long and extends toward
the west to cover also the region supposed to be an X-ray emitting jet. 
The X-ray, optical and mid-IR spatial profiles along these slices 
are presented in Fig.~\ref{f:prof}. The horizontal axis in the {\sl Left panel}
is directed from south to north, and in the {\sl Right panel} from east to west.

This figure shows the coincidence of the positions of the main peak of the 
nebula in the optical, mid-IR and in X-rays. The ACIS-S peak is broader 
than that of the HRC because of the lower spatial resolution of ACIS. The 
same is true for the infrared where we show only the 8 $\mu$m profile. The 
5.8 $\mu$m profile is practically the same, while the 3.6 $\mu$m 
and 4.6 $\mu$m profiles are contaminated by the nearby star east of the 
pulsar. In the {\sl Left panel} the ACIS-S peak has an apparent 
offset toward north by roughly an arcsecond, but this may be related to 
errors in the astrometrical referencing of the ACIS-S image 
\citep{sla02,sla04}.
The profiles confirm that we see the same source in all three spectral domains.
Although the position of the main optical and IR peaks coincide with 
the pulsar X-ray position, we do not resolve any point-like object in the
center of the nebula  in our optical and the archival IR images. 
The FWHM of the optical PSF is $\sim$0\farcs7-0\farcs85 (Sect.~2.1), while for 
the mid-infrared it varies with the wavelength from $\sim$2\farcs1 to 2\farcs5.    
The brightest part of the PWN, which subtends several arcseconds, is clearly 
resolved (Fig.~\ref{f:prof}). To resolve the point-like pulsar from the 
nebula in the optical/IR would require deeper imaging at high spatial 
resolution, or alternatively time resolved observations.

We also note a broad secondary peak west of the pulsar seen in 
the east-west optical and IR profiles and which is marked by a vertical 
dotted line in the {\sl Right panel} plot of Fig.~\ref{f:prof}. 
It coincides with a shallow secondary peak seen in the ACIS profile. This 
is Filament ``A" in Fig.~\ref{f:3c58} with its knots, and which is projected 
on the extended X-ray structure which was early supposed to be the pulsar jet. 

To summarize this part, we conclude that the coincidence of the center 
positions of the detected optical/mid-IR nebula with that of the 
pulsar/PWN X-ray source, the optical/mid-IR/X-ray morphology and the   
spatial surface brightness profiles in these wavelength regions, strongly 
support that we have indeed detected the optical/mid-IR counterpart of 
the pulsar/PWN system in the 3C~58 supernova remnant. 
\subsection{Optical and mid-IR photometry} 
Optical photometry of the suggested  counterpart was performed on the star 
subtracted images. We used the elliptical apertures from the surface brightness
fit described in Sect.~3.2. The parameters of the ellipses are presented in 
Table~\ref{t:ell}. The backgrounds were estimated from a circular annulus with 
a $\sim$20 pixel inner radius and a width of $\sim$10 pixels centered on the
nebula center. The relative photometric errors were minimized ($S/N\approx13$)
for the aperture that formally encapsulates  $\ga86\%$ of the total elliptic nebula flux.
The measured magnitudes of the integral pulsar/torus  nebula emission 
are  $B=24\fm06\pm0.08$ and $V=23\fm11\pm0.04$.
We also tried circular and polygonal apertures of different geometries  
to better encapsulate the whole flux, but got practically the same results. 
The magnitudes were transformed into fluxes 
using  the zero-points  provided by \citet{Fukugita}. 
The results are summarized in Table~\ref{t:phot}. 
The magnitude  distribution with the nebular radius can be estimated using 
these magnitudes and  the normalized fluxes for all elliptical 
apertures which are given in Table~\ref{t:ell}. 
\begin{figure}[t] 
\includegraphics[width=7.0cm,angle=0,clip]{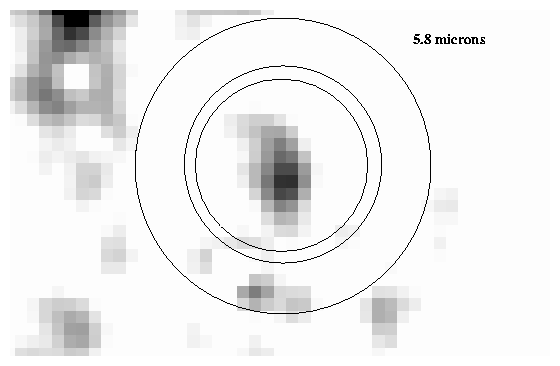}
\caption[h]{Enlarged image of the pulsar/PWN  candidate counterpart at 
5.8 $\mu$m. Nearby background stars have been subtracted and the image was 
smoothed with a 2\farcs4 Gaussian kernel.   
Circles with 8\farcs4, 9\farcs6, and 14\farcs4 radii show the aperture 
and background annuli used for the photometry.               
}
\label{f:spitzer-sub5.8-ima}
\end{figure}    

All nearby background stars were also successfully subtracted from the 
Spitzer/IRAC pipeline post-BCD mid-IR images using the {\sl psf} and {\sl allstar} IRAF utils.
Then aperture photometry with circular apertures was done using 
{\sl daophot} tasks in accordance with  prescriptions given  
in the IRAC Observers Manual\footnote{see, e.g., ssc.spitzer.caltech.edu/archanaly/quickphot.html}. 
A typical aperture radius, where the curves of growth saturate, was about
6--7 pix, or 7\farcs2-8\farcs4 depending on the band, and  the annulus for 
backgrounds was 8--12 pixels. An example of the aperture choice for the 
5.8 $\mu$m image is shown in Fig.~\ref{f:spitzer-sub5.8-ima}. 
 We repeated  the star subtraction procedure 
and varied the background region. The differences in the magnitudes obtained
were about the measurement statistical errors and they 
were included in the resulting uncertainties. 
The magnitudes  were converted into fluxes in physical units and the results are presented 
in Table~\ref{t:phot}. Extended source aperture corrections were applied for 
the flux values\footnote{ssc.spitzer.caltech.edu/irac/calib/extcal/index.html} 
and the flux errors are combined from the magnitude measurement errors 
and $\sim$5\% IRAC calibration uncertainties in each band.  
\begin{table}[t]
\caption{Observed magnitudes and fluxes for the presumed optical/infrared 
pulsar/PWN counterpart of 3C~58, as well as dereddened fluxes for different 
values of $A_V$.}
\begin{tabular}{llllll}
\hline\hline 
  $\lambda_{eff}$          & Mag.       &  log Flux          &                       & log Flux              &         \\ 
  (band)            &  obs.$^a$  &  obs.$^a$          &                       &  dered.$^a$           &         \\  
\cline{4-6}
              &            &                    & $A_V$=1.9             &  2.9                  & 3.4    \\ 
($\mu$m)      & (mag)      & ($\mu$J)           & ($\mu$J)              & ($\mu$J)              & ($\mu$J)\\ 
\hline 
 440($B$)    & 24.06(8)       & $-$0.02(3)         & 0.99(3)               & 1.52(3)               & 1.78(3)  \\ 
 530($V$)    & 23.11(4)       & 0.31(2)            & 1.10(2)               & 1.52(2)               & 1.73(2)  \\ 
 650($R$)$^b$    & $>$22.5(5) & 0.48(20)           & 1.11(20)              & 1.44(20)              & 1.61(20)  \\ 
 3.6         & 15.8(4)        &2.14(16)            & 2.19(16)              & 2.22(16)              & 2.23(16)  \\
 4.5         & 15.2(4)        &2.16(14)            & 2.20(14)              & 2.22(14)              & 2.22(14)  \\
 5.8         & 14.3(3)        &2.34(11)            & 2.38(11)              & 2.40(11)              & 2.41(11)  \\
 8.0         & 13.5(2)        &2.42(9)             & 2.46(9)               & 2.48(9)               & 2.49(9)  \\
\hline
\end{tabular}
\label{t:phot}
\begin{tabular}{ll}
$^a$~numbers in brackets are 1$\sigma$ uncertainties referring to last significant& \\
  digits quoted & \\ 
 $^b$~upper limit for the pulsar magnitude taken from \citet{fes07}   & \\
\end{tabular}
\end{table}

As seen from Table~\ref{t:phot}  the measured mid-IR fluxes  
have larger uncertainties than  the optical ones. 
The main reason is the faintness of the mid-IR source which is 
detected at about the 5$\sigma$ level. The 10\%--15\% uncertainties introduced by  using  
the pipeline produced mosaic IR-images  does not exceed 
the count statistic errors and cannot considerably change the results. 
 Thus, we confirm, but at a higher  level of significance, the 2$\sigma$ mid-IR detection 
reported by \citet{sla08}.   The difference between the mid-IR and  
the optical fluxes is quite significant, and allows us to draw conclusions 
on the multiwavelength spectral energy distribution (SED)  of the suggested counterpart.         

\subsection{Multiwavelength spectrum of the pulsar+PWN source}
We can use our fluxes estimated for the pulsar/PWN  counterpart candidate 
together with the X-ray data to construct a tentative multiwavelength spectrum 
for the central part of the nebula. 
When compiling data from the IR to X-rays one has to take into account that 
the PWN torus size in the optical and IR appears smaller than in the X-rays.
Future deeper optical and IR studies will probably reveal 
fainter outer parts of the torus in these ranges, 
as has been seen for the Crab and PSR B0540-69 PWNe. 
These fainter but more extended outer parts can contribute considerably 
to the total flux of the system.

\begin{figure*}[t]
\setlength{\unitlength}{1mm}
\center{
\includegraphics[width=12.5cm,angle=90, clip]{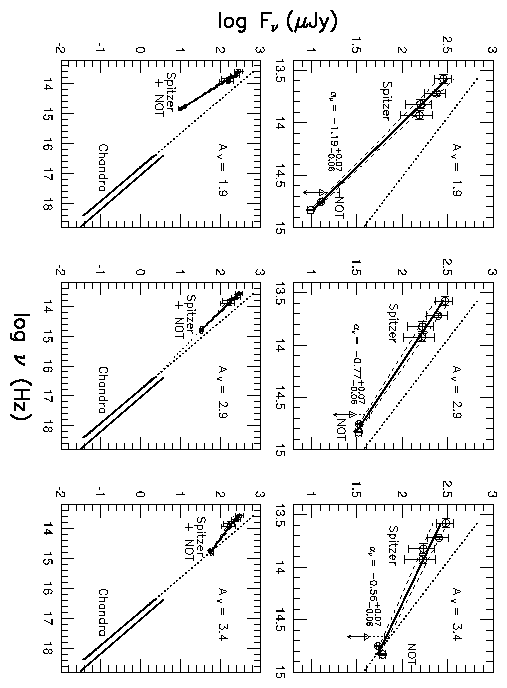}
}
\caption{ Tentative unabsorbed multiwavelength spectrum for the 
inner part of the torus region of the 3C~58 pulsar/PWN system compiled from 
data obtained with different telescopes, as indicated in the plots. 
Different panels demonstrate the dependence of the spectrum on the adopted interstellar 
extinction $A_V$.  The {\sl Lower and Upper panels} show the whole IR/X-ray spectral range and its 
enlarged  optical/IR part, respectively.  Both the optical/IR and X-ray data are fitted 
with power laws with different spectral indices (defined
as $F_{\nu} \propto \nu^{\alpha_{\nu}}$). The derived  indices
for the optical/IR part are shown in the {\sl Upper panels}  and dashed lines 
constrain 1$\sigma$ uncertainties of the respective fits.  For the X-ray part, 
the lower power law is for the inner part of the torus (i.e., same as for
the optical/IR emission), whereas the upper power law is for the entire
torus. The power-law indices for these X-ray data are $-0.88\pm0.08$ 
and $-0.87\pm0.02$, respectively. Dotted lines show extensions 
from the optical/IR part and the X-ray part. Grey polygons (barely resolved)
are 1$\sigma$ uncertainties of the  X-ray fits. 
For the optical/IR range the steepness of an assumed 
power law depends strongly on $A_V$, with an upper limit 
of $\alpha_{\nu} \sim -0.5$. For  $2.9 \leq$ $A_V$ $\leq 3.4$, the 
optical/infrared spectrum can be joined with the X-ray spectrum using only 
one spectral break. 
Lower extinctions suggest a more complicated multiwavelength spectrum. 
The upper limit for the pulsar flux in the $R$ band \citep{fes07} 
is included. Its possible uncertainty (0.5 mag) is marked by a dashed 
errorbar. See text for further details.
}
\label{f:mw}
\end{figure*}

At the current stage it is reasonable to compare the 
measured optical and infrared fluxes with the X-ray spectrum 
extracted from the same physical region. 
To do that we extracted the X-ray spectrum for the central part of the torus 
region from the ACIS-S data making use  of the {\tt CIAO} {\tt acisspec} 
tool and the same elliptic aperture we applied for the optical source 
photometry.     This gave $\sim$30200 source counts. 
The spectral data were grouped to provide a minimum of 20 counts per spectral 
bin and fitted by an absorbed power-law model using standard {\tt XSPEC} 
tools. We obtained a statistically acceptable fit with the photon spectral index 
$\Gamma=1.88\pm0.08$, the absorbing column density $N_{\rm H}=(4.34\pm0.08)\times10^{21}$ cm$^{-2}$, 
and a normalization constant $C=(4.39\pm0.17)\times10^{-4}$~photons~cm$^{-2}$~s$^{-1}$~keV$^{-1}$. 
The fit had $\chi^2=0.95$ per degree of freedom.  Within the uncertainties 
the $\Gamma$ and $N_{\rm H}$ values obtained are in 
agreement with those obtained for the entire torus by \citet{sla04},   
while the unabsorbed integral flux, 1.68$\times10^{-12}$~erg~cm$^{-2}$~s$^{-1}$  (0.5--10 keV range),   
is only $\sim40\%$ of that from the entire X-ray 
torus region.  The derived unabsorbed X-ray spectra for both
the inner  region and the entire torus regions are shown in Fig.~\ref{f:mw}.

Before comparing with the X-ray spectrum we need to correct the optical and 
IR data for interstellar reddening. 
The interstellar color excess towards 3C 58 is, however, not well constrained. \citet{fes83}
estimated, based on the H${\alpha}$/H${\beta}$ decrement obtained from spectral observations  
of a part of the bright long E-W filament discussed in Sect.~3.1, 
that $E(B-V)=0.6\pm0.3$~mag, assuming an intrinsic H${\alpha}$/H${\beta}$ ratio of 3.0. 
This corresponds to the $V$-band reddening correction of $A_V\approx1.9\pm1.0$,   
adopting the standard ratio $A_V/E(B-V)=3.1$. Later spectroscopy of a sample 
of the remnant emission knots provided a more stringent constraint, 
$E(B-V)=0.68\pm0.08$~mag \citep{fes88}.  
Recent  spectroscopy of three brightest H${\alpha}$-filaments in  
the northern part of the SNR yielded a similar 
result,  $E(B-V)=0.6\pm0.1$~mag \citep{fes07}.

We also know that 3C~58 sits within the Galactic disk (l$\approx$130\fdeg4, b$\approx$3\fdeg04) 
and that the entire Galactic excess in this direction provided by \citet{schleg98} 
is $E(B-V)\approx 0.99$~mag. This is consistent with the fact 
that the Galactic disc edge in this direction of the sky is roughly 1--2 kpc more distant 
from us than 3C~58. Assuming a linear color excess increase with the distance from the Sun  
with a  mean gradient of $\sim$0.2~mag~kpc$^{-1}$  applicable within 3--4 kpc 
of the solar neighbourhood  \citep[e.g.,][]{boh1978}, 
we obtain $E(B-V) \sim 0.6$ for 3C~58 and $\sim$1 mag  
as the entire excess, respectively, provided that the distances to 3C~58  and the disc edge 
are 3.2~kpc and $\sim$5~kpc. This is in good agreement with the estimates above.      
      
On the other hand, using  $N_{\rm H}$=$(4.34\pm0.08)\times10^{21}$ cm$^{-2}$, derived 
from the above X-ray spectral fit, and the empirical relation 
$N_{\rm H}/E(B-V) =  4.8\times 10^{21}$~cm$^{-2}$~mag$^{-1}$ 
applicable for the Milky Way \citep{boh1978}, we obtain
$E(B-V)=0.9\pm0.02$~mag ($A_V=2.8\pm0.06$). 
Another empirical relation between the effective $N_{\rm H}$ of the X-ray absorbing 
gas and the dust extinction, $N_{\rm H}/E(B-V) = (5.55\pm0.093)\times 10^{21}$~cm$^{-2}$~mag$^{-1}$ \citep{pred95}, 
gives a smaller value, $E(B-V)=0.78\pm0.03$~mag.
The color excesses  based on the $N_{\rm H}$ value,       
derived from the X-ray spectral fit of the  torus emission,  
are systematically higher than the mean excess obtained  
from the optical studies. This may be explained by intrinsic reddening  variations 
over the  remnant area resulted from a nonuniform  
distribution of the SN ejecta.   

Summarizing all the uncertainties of the color excess 
we select for our further analysis $A_V=1.9$~mag 
as a most plausible reddening consistent with 
the mean excess given by the optical studies.
We also perform the analysis for 
$A_V=2.9$ which is compatible with the  X-ray spectral estimates discussed above.  
Dereddened optical/IR fluxes 
are presented in Table~\ref{t:phot}. 
For dereddening  we used a standard optical extinction curve \citep{card89} 
and  average $A_{\lambda}/A_K$ ratios  provided especially for the Spitzer/IRAC bands 
by \citet{indeb05}. 
 
We have combined dereddened optical/IR fluxes for the pulsar/PWN system with 
the unabsorbed X-ray spectrum in Fig.~\ref{f:mw}. As expected, the shape 
of the multiwavelength spectrum strongly depends on $A_V$. 
To emphasize this we have also included an even higher value for the extinction towards
3C~58, with $A_V$=3.4 mag. This value corresponds to the upper limit of 
Galactic extinction in the given direction  as 
derived from the total neutral hydrogen Galactic 
column density,  $N_{\rm H}=5.34\times 10^{21}$~cm$^{-2}$ \citep{kal05}, making use 
of the relation by \citet{boh1978}. This higher value of extinction is included becasue 
it represents an interesting limiting case for the dereddened spectral energy distribution (see below). 
In Fig.~\ref{f:mw} we have plotted the power-law describing the X-ray 
spectrum from the entire PWN torus (upper X-ray power law). 
We also show the X-ray spectrum from the only same physical region  
as for the optical/IR emission (lower X-ray power law).

In all cases  the optical/IR SED  can be  fitted 
by a power law ($F_{\nu} \propto \nu^{\alpha_{\nu}}$) and the spectral 
indices, $\alpha_{\nu}$, of the fits are shown in the upper panels
of Fig.~\ref{f:mw}. For the most plausible $A_V$ value of 1.9  the fit 
residuals are minimal and we see a monotonous decline in flux from the IR 
to the optical, suggesting a nonthermal nature of the detected  
optical/IR counterpart candidate. This implies a synchrotron emission 
mechanism for the detected nebula, normally considered to be the main 
radiative process for PWNe. For higher extinction values, particularly 
for $A_V=3.4$, the spectrum becomes  flatter.

The upper limit of the pulsar point source flux in the $R$-band obtained 
by \citet{fes07} is also shown in the plots for comparison. It was not included 
in the fits and we estimate that its uncertainty 
can be as large as half a stellar magnitude. This is becasue it was derived 
from a comparison with  rather uncertain catalogue magnitudes of 
USNO stars in the 3C~58 field.   
We included  the possible flux uncertainty in Table~\ref{t:phot} and 
Fig.~\ref{f:mw}. The pulsar $R$-band flux upper limit is a factor of a few 
lower than the expected nebula flux in this band, which is compatible with 
our estimates for the pulsar flux contributions 
in the $B$ and $V$ bands (see Sect. 3.2).   
 
Extending the X-ray spectral fits ({\sl dotted lines}) toward the optical/IR 
bands, and {\sl vice versa},  we  see that  at any adopted $A_V$ the X-ray 
extension overshoots the IR fluxes. For the $A_V$ range of $\sim$2.9 --3.4 the 
multiwavelength spectrum from the IR through X-rays can be modeled by a power
law with a single break and a flatter slope in the IR/optical range. In the 
optical/IR range the spectrum  cannot be shallower than that 
with $\alpha_{\nu}\sim -0.5$, as follows from the largest value 
of $A_V = 3.4$ ({\sl Right panels}), when the break occurs directly in 
the $B$ optical band. For lower $A_V$ the break moves toward the X-ray range 
and at $A_V = 2.9$, it meets the low-energy boundary of the X-ray range  ({\sl Middle panels}). 
In this case the optical/IR slope with $\alpha_{\nu} \approx -0.8$  is close to 
that in X-rays with  $\alpha_{\nu} \approx -0.9$. For lower values of $A_V$ 
the intrinsic optical/IR spectrum becomes steeper and we must invoke more 
breaks to connect the optical/IR SED with the X-ray spectrum.        
For example, at $A_V = 1.9$, the optical/IR spectral index, $\alpha_{\nu} \approx -1.2$, 
is significantly lower than that in X-rays  ({\sl Left panels}).

The Crab and  PSR B0540-69 are the only two PWNe previously 
detected in the optical and mid-IR ranges. 
Even within the large uncertainties of the interstellar extinction toward 3C~58, 
the tentative multiwavelength spectrum of the 3C~58 PWN  can be similar to either
the Crab PWN (for $A_V \gsim 2.9$), or the PWN of PSR B0540-69 (for $A_V \lsim 2.9$). 
The Crab PWN changes smoothly its spectral slope from 
negative in X-rays with $\alpha_{\nu}$$\sim-1$ to a flatter value in the 
optical, whereas the spectrum of the PSR B0540-69 PWN demonstrates a 
double-knee structure at the transition from the optical to X-rays \citep[e.g.,][]{ser04}.
Our preferred value of $A_V \sim 1.9$ suggests that the 3C~58 PWN is likely similar to the latter case.  

We emphasize the importance  of using the same spatial region in all
spectral domains when constructing the multiwavelength spectrum. 
As seen from the {\sl Bottom panels} of Fig.~\ref{f:mw}, using the entire torus 
X-ray spectrum, instead of its inner region, would lead to different  
conclusions. To make a meaningful multiwavelength plot for the entire 
X-ray emitting PWN torus, we would need deeper optical/IR observations to 
constrain the emission in the outer parts of the torus, and to make a more 
detailed comparison with the entire PWNe of the Crab and PSR B0540-69 and with available theoretical models.
\section{Discussion}
The properties of the detected source indeed suggest 
it to be the optical/mid-IR counterpart  of the 
3C~58 pulsar/PWN torus system. 
The alternative could be a SNR filament, or possibly a faint 
background spiral galaxy seen edge on and coinciding by chance with 
the pulsar position. However, the positional coincidence with the X-ray 
PWN torus and the remarkable similarity of the object morphology and 
orientation with those in X-rays is reassuring. The multiwavelenght spectrum 
shows that the optical and mid-infrared magnitudes of the proposed 
counterpart are at least consistent with the spectra seen in other PWNe.
We argue that this makes the alternative interpretations rather unlikely.

There are no significant radio filaments at the PWN position  
\citep{beit07} 
The nearest bright and sharp long filament known  
as a ``wisp"  \citep{frail93} 
is a few arcseconds west of the 
PWN torus and the optical/IR source boundaries. Some signs of a faint 
radio counterpart of the PWN torus and/or the optical source may be present 
in the latest  VLA 1.4~GHz image 
published by \citet[][Fig.~2]{beit07}.

For an $A_V$ value consistent with the $N_H$ column density derived from the X-ray 
spectral fit there are $\sim 1\sigma$ deviations of the B and V fluxes 
from the best power law fit line describing the optical and mid-IR SED 
of the suggested counterpart (see {\sl Middle
panels} of Fig.~7). Additional optical observations are in progress to extend 
the SED toward the red and UV optical bands and to see if these 
deviations are significant and if they can better constrain the extinction 
value, assuming a single power-law fit of the optical/IR SED. 
Higher spatial and/or timing resolution are necessary to measure 
reliably the optical emission from the point-like pulsar embedded in the 
center of the  PWN torus and  to confirm finally the suggested identification.  
Longer wavelength mid-infrared imaging with the Spitzer/MIPS is needed to see
if there is a break toward the radio to support the apparent absence/presence of 
the putative PWN torus counterpart in the radio range. 

At larger spatial scales the spectrum from the whole 3C~58 plerion region 
from radio through X-rays is described by a power law with a single low 
frequency break near $\sim$50 GHz. Below this frequency the spectral slope 
is significantly flatter than in X-rays \citep[cf.][]{sla08}.  
This is in contrast to the plerion core near the pulsar. For the PWN torus 
our optical data suggest at least one break at much shorter wavelenghts, 
somewhere between the optical and X-ray ranges. This implies a different 
spectrum of the relativistic particles in the plerion core at distances  
from the pulsar comparable to the pulsar wind termination shock radius where 
the PWN torus structure is formed. 
At any reasonable assumption on the interstellar extinction  value,  
the optical/IR part of the power-law torus spectrum cannot be flatter than  
that with  $\alpha_{\nu} = -0.5$ and steeper than that with  $\alpha_{\nu} = -1.2$.  
Comparison with the radio flux from the torus or its upper limit 
would provide additional constraints of the spectrum of the emitting particles.     
\begin{table*}[t] 
\caption{Comparison of the optical and X-ray spectral indices 
(${\rm \alpha^{O}_{\nu}}$, ${\alpha^{\rm X}_{\nu}}$) luminosities 
(${L^{\rm O}}$, ${L^{\rm X}}$), efficiencies (${\rm \eta ^{O}}$, 
${\rm \eta ^{X}}$) of the three young PWNe detected in the optical/IR 
and X-rays. Information on the Vela PWN, not yet detected in the optical, 
as well as on the pulsar characteristic ages ($\tau$), spindown pulsar 
luminosities (${\dot E}$), PWN sizes, and the ratios of the pulsar to PWN 
luminosity in the optical and X-rays are also included.    
}  
\label{t:pwn-lum}
\begin{center}
\begin{tabular}{lccccccccccc}
\hline\hline 
PWN & $\tau$ & ${\dot E}$ & size & ${\rm -\alpha^{O}_{\nu}}$ & ${L^{\rm O,a}}$ & ${\rm \eta ^{O}}$ & ${-\alpha^{\rm X}_{\nu}}$ 
& ${L^{\rm X,b}}$ & ${\rm \eta ^{X}}$ 
& ${L^{\rm O}_{\rm psr}/L^{\rm O}_{\rm pwn}}$ & ${L^{\rm X}_{\rm psr}/L^{\rm X}_{\rm pwn}}$ \\  
  & kyr  & ${\rm 10^{37}erg~s^{-1}} $  &pc & & ${\rm 10^{33}erg~s^{-1}}$ & ${\rm 10^{-5}}$ &  & ${\rm 10^{36}erg~s^{-1}}$ & ${\rm 10^{-3}}$ 
    &  & \\
\hline  
Crab$^c$  &1.24 &46     &1.5        & 0.92 & 4240   & 920        & 1.14  & 21.8                     & 47.5                       &0.0017    &0.046 
\widerul \\
0540$^c$  &1.66&50.2   &0.6-0.9   & 1.5  & 366    & 245        &  1.04  & 12                       & 79.7                       &0.03      &0.26 
\widerul  \\    
3C 58$^d$ &5.38&2.6    &0.08-0.19 &0.7-1.2  & 0.08-0.21 & 0.3-0.75 & 0.88       & $5\times 10^{-3}$            & 0.19                       &$\la0.1$  & 0.23 
\widerul  \\ 
Vela$^c$  &11&0.069  & 0.14      & --       & --         & --         & 0.5    & ${\rm 6.8\times 10^{-4}}$  & ${\rm 9.8\times 10^{-2}}$  & --       & 0.34
\widerul  \\
   
\hline
\end{tabular}
\label{t:pwn-params}
\begin{tabular}{ll} 
$^a$~For the optical range 1.57--3.68 eV. & $^c$~Data are taken from \citet{ser04}. \\
$^b$~For the X-ray range 0.6--10 keV.     & $^d$~This paper. The distance is assumed to be 3.2 kpc.   \\
\end{tabular} 
\end{center}
\end{table*} 

Assuming that we have indeed detected the optical/IR counterpart of the 
torus-shaped PWN, we compare in Table~\ref{t:pwn-params} its parameters 
with the parameters of the other two young pulsar/PWN systems which have been 
detected in the optical, infrared and X-rays. 
We also include the  older Vela pulsar whose PWN has not yet been identified 
in the optical or IR \citep[e.g.,][]{shib03}. For 3C~58 we adopt a 
distance of 3.2 kpc \citep{sla04}. Its optical luminosity uncertainty is 
mainly due to the uncertainty in the interstellar extinction.  
The ratio of the pulsar to PWN optical luminosity reflects the possible 
contribution of the pulsar  to the total pulsar/PWN optical emission, as was 
estimated  in Sect.~3.2. The principal ranking in Table~\ref{t:pwn-params} is the characteristic age 
and/or spindown luminosity ${\dot E}$. 
The age of 3C~58 and its association with the historical supernova 
event of 1181~A.D.~is still debated \citep[e.g.,][]{sla04}.
We have used the characteristic age of the pulsar defined in the standard way 
as $\tau =P/2{\dot P}$~\citep{mu02}. 
We see that  the 3C~58 PWN nicely fits  its position in Table~\ref{t:pwn-params}  
in terms of other physical parameters.  
Its size, X-ray luminosity and efficiency of transformation of the pulsar 
rotational loss to PWN emission, $\eta^{X}=L^{X}/{\dot E}$, and the ratio of 
pulsar to PWN X-ray luminosity trace the general evolution tendency 
where the ``strength" of a PWN correlates with the spindown luminosity and 
fades with pulsar age. Interestingly, the luminosity of 3C~58 has faded an 
order of a magnitude more in the optical than in X-rays, if we take the Crab 
PWN as a reference. For PSR~0540-69 and 3C~58 both luminosities have decreased
by approximately the same factor.  This perhaps means that the old estimates of the Crab PWN optical 
continuum luminosity based on low spatial resolution observations
 \citet{VW93} need a revision using recently available high resolution observations of this PWN.  
There is no strong difference for the efficiency 
of the 3C~58 PWN in the optical  and X-rays. The same situation applies to
the younger  PSR~0540-69 PWN, although 3C~58 is much less efficient, as 
expected from its age. For the Crab, the difference is exceptionally strong, 
probably for the same reasons we have mentioned  above. 
Thus, the properties of the likely optical/IR counterpart of the 3C~58 PWN 
suggest that the real age of the associated SNR is close to the pulsar 
characteristic age, but not to the historical event of 
1181~A.D, as has been discussed from other points of view by 
\citet{chev04}  and \citet{beit07}.

The long extended structure seen in X-rays west of the 3C~58 pulsar has been 
suggested to be the jet part of its PWN \citep{sla04}. A similar 
axisymmetrical jet structure of the Crab PWN has a clear optical counterpart
\citep[e.g.][]{hest02}. Some hints of a less pronounced X-ray jet of the 
PSR~0540-69 PWN are also seen in the optical \citep{ser04}. 
For 3C~58 we also see a bright optical filament projected on the X-ray jet
where it bends southward. However, a comparison of the optical, infrared and 
radio images suggests that this is not an optical jet counterpart. It may 
also be that this distant part of the X-ray structure does not relate to the
PWN jet at all. Instead, it is most likely a long curved SNR filament, which 
is also clearly seen in VLA images \citep{beit07}.   
Its position is marked in the 8 $\mu$m images of Fig.~\ref{f:B-V-hrc-ac-sp-vla}. 
Most likely, the part of the filament we have detected in our images, is the same 
as was studied spectroscopically  by \citet{fes83}
(position 1 in his notation) using a 4\asec\ circle aperture. 
Based on H$\alpha$, he estimated a radial velocity of $\sim 900$~km~s$^{-1}$,
which is similar to the overall expansion of the remnant, and to motions
in the Crab. Detailed imaging and spectral studies are necessary to confirm 
this and to study if this extended structure is a part of the real PWN jet or not. 
Deeper spectral observations will also set more stringent constraints on 
the interstellar extinction, which is the main source of uncertainty of 
the estimated intrinsic optical flux of the PWN. 

In conclusion, our observations have, with high probability, increased the number of the optically 
identified Crab-like PWNe from two to three, which should help to constrain 
models of these unique objects and to understand their nature. 
  
\begin{acknowledgements}
The work was partially supported by RFBR (grants 05-02-16245, 05-02-22003),
the Swedish Research Council and Nsh 9879.2006.2. The Dark Cosmology Centre is 
funded by the Danish National Research Foundation. The projected started
while PL was still a Research Fellow at the Royal Swedish Academy, supported 
by a grant from the Wallenberg Foundation. The data presented here 
have been taken using ALFOSC, which is owned by the Instituto de Astrofisica 
de Andalucia (IAA) and operated at the Nordic Optical Telescope.
\end{acknowledgements}

\end{document}